# A possible YORP effect on C and S Main Belt Asteroids

A. Carbognani

*Astronomical Observatory of the Autonomous Region of the Aosta Valley (OAVdA)*

*Saint-Barthélemy Loc. Lignan, 39*

*11020 Nus, Aosta – Italy*

Number of manuscript pages: 29

Number of figures: 6

Number of tables: 3

Editorial correspondence to:

Dr Albino Carbognani

Astronomical Observatory of the Autonomous Region of the Aosta Valley (OAVdA)

Fondazione Clément Fillietroz-ONLUS

Saint-Barthélemy Loc. Lignan, 39

11020 Nus, Aosta – Italy

Telephone: +39 0165 77 00 50

Fax: +39 0165 77 00 51

E-mail address: albino.carbognani@alice.it

**Abstract**

A rotating frequency analysis in a previous paper, showed that two samples of C and S-type asteroids belonging to the Main Belt, but not to any families, present two different values for the transition diameter to a Maxwellian distribution of the rotation frequency, respectively 48 and 33 km. In this paper, after a more detailed statistical analysis, aiming to verify that the result is physically relevant, we found a better estimate for the transition diameter, respectively $D_C = 44 \pm 2$ km and $D_S = 30 \pm 1$ km. The ratio between these estimated transition diameters, $D_C/D_S = 1.5 \pm 0.1$, can be supported with the help of the YORP (Yarkovsky-O'Keefe-Radzievskii-Paddack) effect, although other physical causes can not be completely ruled out.

In this paper we have derived a simple scaling law for YORP which, taking into account the different average heliocentric distance, the bulk density, the albedo and the asteroid "asymmetry surface factor", has enabled us to reasonably justify the ratio between the diameters transition of C-type and S-type asteroids. The same scaling law can be used to estimate a new ratio between the bulk densities of S and C asteroids samples (giving $\rho_S/\rho_C \approx 2.9 \pm 0.3$), and can explain why the asteroids near the transition diameter have about the same absolute magnitude. For C-type asteroids, using the found density ratio and other estimates of S-type density, it is also possible to estimate an average bulk density equal to $0.9 \pm 0.1$ g cm$^{-3}$, a value compatible with icy composition. The suggested explanation for the difference of the transition diameters is a plausible hypothesis, consistent with the data, but it needs to be studied more in depth with further observations.

## 1 Introduction

In a previous paper (Carbognani, 2010), hereafter paper I, we have compared the observable properties of 962 numbered MBAs (Main Belt Asteroids) of Tholen/SMASSII C and S class (495 S and 467 C), with a diameter range 1-500 km, not belonging to families or binary systems. The data for the MBAs sample were drawn from Asteroid LightCurve Data Base (or ALCDB, version of 2009 April 21), care of Alan W. Harris and Brian D. Warner of Space Science Institute and Petr Pravec of the Astronomical Institute, Czech Republic (Warner et al., 2009).

One of the minor results of the paper quoted above is that, between 48 and 200 km, the C asteroids have a rotation frequency distribution compatible with a Maxwellian whereas, for S asteroids the compatibility with the Maxwellian involves diameters greater than 33 km. The transition diameter from non-Maxwellian to Maxwellian appears sharp and one of the questions we want to answer here, is if the diameter transition ratio, i.e. $D_C/D_S$, is relevant from a physical point of view.

Before answering this and other questions, we will review the known physical characteristics considered here for the two asteroids populations: geometric albedo, mean semimajor axis and bulk density.

The C and S class asteroids can effortlessly be distinguished in a (B-V)-(U-B) diagram, and appear as two homogeneous body populations with different surface physical properties (at least in first approximation). The asteroids of Tholen S class have an average geometrical albedo $\bar{p}_v = 0.20 \pm 0.07$, while the C class have $\bar{p}_v = 0.06 \pm 0.02$ for the same quantity (Shevchenko and Lupishko, 1998). These values are practically independent of the diameter.

The majority of MBAs have semimajor axes within the range 2.2-3.2 AU, but the S-type asteroids are located nearer the inner edge of the Main Belt, while the C-type are nearer the outer edge. The dichotomy in the mean distance from the Sun is visible in the two analysed samples. For the S asteroids of our sample the mean heliocentric distance is 2.538 ± 0.011 AU (mean value of 495 objects), while for C the mean heliocentric distance is 3.013 ± 0.014 AU (mean value of 467 asteroids). The semimajor axis data for the MBAs samples are drawn from the JPL Small-Body Database Browser (JPLSBDB), available from http://ssd.jpl.nasa.gov/sbdb.cgi. The distributions are shown in Fig. 1.

The asteroid's bulk density is a more difficult parameter to obtain, as we need to know both the body mass and the volume. Data on the bulk density of asteroids have increased in the last years and have led to significant insights into the structure of these objects. The advancement of knowledge comes mainly from observations of asteroid satellites, asteroid mutual gravitational events, perturbations on neighbouring spacecraft and dynamical models of perturbing acceleration on Mars. In general, most asteroids have relatively low mean densities and probably have significant porosity (Britt et al., 2002). Furthermore there are clues suggesting a density variation within taxonomic classes due to porosity variations (Baer and Chesley, 2008). In our case, we assume that the bulk density is diameter independent. Dynamical estimates of the bulk density of the 300/351 largest asteroids of the Main Belt, plus a massive asteroid ring to take into account the collective gravitational effect of small asteroids, leads to the values shown in Table 1 (Krasinsky et al., 2002). The density estimate shown in the table was made as follows.

A first estimate of the mass of the largest asteroid is made using an estimate of the diameter and the bulk density value for the taxonomic class. Accumulating the planetary perturbations exerted by each class, it is possible to correct the value initially assumed for  bulk density, and determine the best average value representative of the whole taxonomic class. As we can see, the values of the density ratio from Table 1 are not consistent with each other and this gives a glimpse of the difficulty to accurately estimate  asteroids' bulk density. In the following paragraph we will give a quick guide to the paper.

After a review of the method used to calculate the transition diameter and a control of the asteroids database, we have shown that the ratio between the transition diameters is relevant from a physical point of view and that this may be due to YORP (Yarkovsky-O'Keefe-Radzievskii-Paddack) effect on the two different asteroids samples although, at this stage, other physical causes can not be completely ruled out.

To this purpose, we have developed a simple physical model, YORP based, which gives the ratio between the transition diameters as a function of asteroids bulk density, mean heliocentric distance and a quantity that we call "asymmetry surface factor" (see below).

As a first raw test for the physical model proposed here, i.e. to see if the theoretical ratio between the transition diameters is near, or at least not too far, with the one observed, we need to know a mean ratio between the bulk density of S and C asteroids, so we use the mean value obtained from Table 1, about $2.0 \pm 0.4$. After this plausibility test, and using only the observed transition diameters ratio, we will see the physical consequences of this model, and provide a new estimate of the bulk density ratio between the S and C asteroids samples. For C-type asteroids, using the found density ratio and a mean of the S-type density from Table 1, it is also possible to estimate an average bulk density equal to $0.9 \pm 0.1$ g cm$^{-3}$. This value appears to be interesting because it suggests the existence of a high percentage of icy asteroids within the C-type. Finally, we have proved  that our model can justify the fact that the absolute magnitude of the asteroids near the transition diameter is approximately the same.

## 2 A closer estimate of the transition diameters to the Maxwellian distribution

As in Warner et al. (2009), "It is important to remember that the data in ALCDB are not the result of a single, well-controlled survey, but are, instead, the compilation of a number of observing programs, each with its own goals and observing parameters". The current lightcurves numbers in ALCDB are over 3700, with about 3100 asteroids that have reasonably determined rotation periods as of December 2008. For MBAs there is a bias against asteroids in the diameter range 3.5-14 km, because they are too faint to be easily worked upon with small instruments (Warner et al., 2009). In paper I, we have found that "for diameter values equal to or greater than about 20 km, the diameter distribution of C and S asteroids [in ALCDB] is about the same because the mean values and standard deviations are similar. Also, the ratio between C and S numbers is almost constant, with values between 3 and 2.7. On the contrary, below 20 km the average diameter of S is much smaller than that of C and the ratio C/S drops to near 1. This difference can be attributed to an observational bias in favour of the small S asteroids". So, the bias on C and S asteroids periods starts around 15-20 km in diameter (with absolute magnitude in the range 12-11), well below the transition diameters that we want to analyse. Regarding the taxonomy in ALCDB, when available, taxonomic classes are taken from the *Asteroids II* database, SMASS II (Small Main-Belt Asteroid Spectroscopic Survey), and from SDSS (Sloan Digital Sky Survey) colors (Warner et al., 2009). As discussed below, we checked the taxonomic classification for asteroids around the transition diameter and removed the uncertain cases (see Table 2).

Now, we are going to briefly review the method used in paper I (refer to it for more details), for the derivation of the transition diameters to the Maxwellian distribution. The analysed diameters, range from 4 km (lower limit) to 200 km (upper limit) and we use Kuiper test (Press et al., 1992), to evaluate the compatibility with a Maxwellian distribution. We gradually increase the inferior diameter value (keeping constant the superior diameter value), by 1 km steps, and, for every step, we execut Kuiper test for the Maxwellian distribution. The compatibility test is considered successful if the confidence level of rejection is equal to or less than 95%. In our case, as already mentioned, the feedback was negative until we reached 33 km for S-type and 48 km for C-type asteroids.

For a direct comparison between the rotation frequency distribution and the three-dimensional theoretical Maxwellian distribution, see Fig. 2, 3, 4 and 5. In these figures, the rotation frequency is indicated by the $\Omega$ symbol and is equal to 24/P (where P is the asteroid rotation period in hours), and it is measured in rotations per day (or cyc day$^{-1}$). As we can see, both for the S and C-type asteroids, the rotation frequencies distribution is in fairly good agreement with the Maxwellian distribution (as confirmed by Kuiper test), only in case the diameter value exceeds a certain threshold. If we go below this threshold, an excess of slow rotators appears, as well as a lack of medium rotators and an excess of fast rotators, which is due to smaller asteroids. In these figures we did not go below 15 km in diameter due to the observational bias.

After that, in order to test any physical relevance of the results found in paper I, we have analysed in depth the accuracy of asteroid periods and diameters around the Maxwellian transition diameter so, for S-type, the examined asteroids have diameters in the range 20-50 km (92 objects), while for C-type the range is 30-60 km (183 objects). To select the asteroids whose rotation frequency deviates too much from the Maxwellian distribution we have used the $\chi^2$ test with a bin of 0.5 d$^{-1}$. We have decided to review the asteroids data for which:

$$\frac{|O - E|}{\sqrt{E}} \geq 1 \qquad\qquad\qquad (1)$$

In Equation (1), $O$ is the observed asteroid number in the bin, while $E$ is the expected asteroid number in accordance with the Maxwellian distribution. From Eq. 1, we have checked the data of 69 S-type asteroids and 150 C-type asteroids with the new version files of ALCDB (up-to-date, November 18, 2009). For a cross comparison we have used the taxonomic and albedo data contained in JPLSBDB. As far as the C and S diameters and the rotation periods values are concerned, there are no changes in the new ALCDB with respect to the values previously used in paper I. For most asteroids (about 60%), the lightcurve quality code is U=3, whereas for the remaining objects it is U=2. The value U=3, denotes a certain result with no ambiguity and full lightcurve coverage while, if U=2, the period is based on less than full coverage, so it might be wrong by 20 percent or more, but still useful for statistical study. For more details on the assignment of the U code see Warner et al., (2009) or the "README.txt" file that accompanies the ALCDB.

Regarding taxonomy, most S-type asteroids were classified according to the spectrum or the color indices. Twelve asteroids are classifiable S-type according to the IRAS (Infra Red Astronomical Satellite) geometric albedo ($p_v = 0.11$-$0.32$); four asteroids have a low IRAS geometric albedo and appear more C-type asteroids ($p_v = 0.047$-$0.056$). For these asteroids there are no other useful data. Finally, asteroid (723) is classified as C-type from JPLSBDB (high resolution spectrum), while asteroid (1407) is classified as X-type (same source and motivation). Thus, in first approximation, for S-type we can discard six asteroids, about 9% of the total (see Table 2 for the full list).

Regarding the C-type asteroids, only 25 were classified by means of spectrum or colour indices. About half are classifiable as C-type owing to the IRAS geometric albedo ($p_v = 0.021$-$0.092$); four asteroids are classified X-type from JPLSBDB (high resolution spectrum); eleven asteroids have a high IRAS geometric albedo ($p_v$ = 0.12-0.27), which makes it unlikely that they belong to C-type. Finally, for twenty-seven asteroids, there are no useful data for taxonomic classification of any kind. So, concerning C-type asteroids, and always as a first approximation, there are 42 objects to be discarded, about 28% of the total (see Table 2 for the list). In general, after this brief analysis, one can see that S-type asteroids are best observed from the physical point of view.

However, after this check and screening of the asteroids database around the raw transition diameters found in paper I, there is an additional problem one has to deal with an additional issue: asteroids' diameters are not exactly known.

In ALCDB the diameters are taken, when available, from the Supplemental IRAS Minor Planet Survey (Tedesco et al., 2002), in short SIMPS. If this value is not available, the diameter is estimated by Eq. (14), with the $H$ value taken from MPCORB catalogue (Minor Planet Center, 2010) and with an assumed value for the geometrical albedo. So, asteroids' diameter values in ALCDB could be incorrect by ± 20% or more (Warner et al., 2009). As for our full samples of C and S examined in paper I, the fractions of asteroids with diameters from SIMPS vs. diameter lower limit are shown in Table 3. As we can see, for asteroids greater than 10 km, most of the diameter values come from SIMPS. For our subsample of S-type asteroids near transition diameter, 94% of the diameter values are from SIMPS, while for C-type asteroids the percentage drops to 73%. In practice, most of our asteroid diameters come from IRAS catalogue.

For asteroids with 100 km < D < 350 km for which there exist high-quality results from stellar occultations, the mean difference in the diameters derived from IRAS is ± 7% (Tedesco et al., 2002), that which rises to ± 35% for small asteroids (Krasinsky et al., 2002). So, regarding mean diameters uncertainty for our asteroids sample, we can assume a random value of ± 20%, without appreciable systematic errors.

A similar argument applies to the rotation frequencies with quality code U = 2. In the S asteroids sample with diameter D ≥ 20 km, these objects amount to 32% of the total while, for C asteroids, the percentage rises to 46%.

Now, the problem is to estimate the mean uncertainty for the periods with lightcurve quality U = 2. To this purpose, we compared the rotation periods of forty asteroids, both C and S, with quality U = 3 and belonging to our sample, with the corresponding period values included in ALCDB of 2006 March 14, when their lightcurve quality was U = 2. The mean relative period uncertainty is 10% ± 3%  so, to be sure not to underestimate  uncertainties, a mean representative value of ± 20% was taken also for periods, as for  diameters.

We do not know the real diameters of asteroids or their rotation frequencies (for the cases U = 2), but now we can estimate the sensitivity and the uncertainty of the transition diameters by adding a gaussian random noise of ± 20% on the known diameters and rotation frequencies of the full S and C asteroids samples, and then the Kuiper test is repeated in the same way as in paper I. As in previous work (paper I), the tests were done normalizing every rotation frequency of the samples to the mean rotation frequency, in order to eliminate the dishomogeneity due to different asteroid sizes (Pravec and Harris, 2000), but there are not significant differences between the results with and without normalization. All this is done for $n$ times (with $n$ sufficiently large, see Appendix). At the computation end, with a set of independent transition diameter values, we can estimate the mean value and the standard deviation.

Applying the method outlined above, the transition diameters appear to be: $D_C = 44 \pm 2$ km and $D_S = 30 \pm 1$ km (see Appendix). These values are in agreement with the first row estimate obtained in paper I (so Fig. 2, 3, 4 and 5 do not change substantially). The transition diameter appears to be not dramatically sensitive to asteroids diameters and rotation frequencies uncertainties. However, uncertainties on the rotation frequencies have a much greater importance than the ones on the diameters. The transition diameter difference is $D_C - D_S = 14 \pm 3$ km, and can be considered a significant result. So, our best estimate for the transition diameters ratio is $D_C/D_S \approx 1.5 \pm 0.1$.

It is interesting to note that rotational analysis carried out considering all asteroid types, indicate the 30-40 km range below which we have large deviations from the Maxwellian distribution, due to the occurrence of several relatively rapidly rotating asteroids and one very slow rotator (Pravec et al., 2002). This transition range appears compatible with our previous values, obtained by analyzing separately the C and S asteroid types.

As stated in Farinella et al. (1981), a difference in the spin rate distribution of two asteroid samples of different taxonomic types can be due to "dynamical" or "chemical-physical" reasons. In the former category we can have different mean sizes of the parent bodies or different mean relative velocities, while in the latter we have difference in structure or composition, resulting in different behaviour during collisional events. But there may be a third category, involving environmental factors external to the asteroid belt, such as solar radiation.

As mentioned above, the distributions of the rotation frequencies in Fig. 2 and 4, in comparison with Fig. 3 and 5, show an excess of fast and slow rotators of little diameter, while the central part of the distribution appears empty, as if an external factor had "smeared" the rotation frequency with respect to Maxwellian distribution. This appears to be independent of taxonomic type and is in agreement with the YORP effect behavior because, if we reduce the asteroid's size, we expect a major alteration of the rotation, due to the thermal emission, toward shorter and long periods (Pravec et al., 2008).

 Now we shall try to explain the *Dc/Ds* value (and more), with the help of this physical effect, but it is important to recall that deviations of asteroid spin rates from the Maxwellian distribution can occur not only when the asteroid population is not collisionally relaxed, but also when the asteroid sample is dishomogeneous (Pravec et al., 2002).

Despite every possible effort to have homogeneous asteroids samples, it may be not so. Therefore, what follows should be considered with caution, as we can not rule out other physical causes for the deviations from Maxwellian distribution. We assume here that YORP is the only, or the main, process affecting asteroids' spin.

## 3   A scaling law for the YORP effect

The YORP effect is a non-gravitational force due to the thermal photon emission from the asteroid surface, and can change the spin of the body if it has the right shape. An asteroid to be spun up must have a certain amount of asymmetry in its shape; figures of revolution as the sphere, or even the triaxial ellipsoid, will not be affected by the YORP effect (Rubincam, 2000). Unlike the photons emitted, the momentum deposited on the surface by arriving solar photons cancel identically when averaged around an orbit, while the recoil imparted to the surface by departing reflected photons is proportional to Bond albedo, and is irrelevant for a dark surface like that of the asteroids, which reflects  sunlight very little (Statler, 2009).

The YORP effect can have a significant effect on the spin even for relatively large asteroids. With a simple scaling of the Rubincam's work, (2000), we can compare YORP vs collision timescale, for C and S asteroids near the transition diameter found above. For C-type asteroids (pseudo-Deimos model), near 3 AU, with a diameter of 44 km and a bulk density of 1.3 g cm$^{-3}$ the time to double the rotation speed is $T_{YORP} \approx 1.4 \cdot 10^9$ years, whereas typically the time to substantially change the rotation period of a Main Belt asteroid is $T_{Coll} \approx 5 \cdot 10^8$ years (Farinella et al., 1998). So, the YORP timescale is longer than the collision timescale but an alteration in the rotation period for C-type asteroids appears still possible. For S-type asteroids (pseudo-Ida model), near 2.86 AU from the Sun, with a diameter of 30 km ad a bulk density of 2.6 g cm$^{-3}$ we have $T_{YORP} \approx 1.9 \cdot 10^8$ years, while $T_{Coll} \approx 4.1 \cdot 10^8$ years so in this case the timescales are fully comparable. Thus the YORP effect, despite collisions, appears physically significant for MBAs spin rates evolution up to bodies of about 40 km in diameter with average density of 2-3 g cm$^{-3}$ (Binzel, 2003).

Moreover, the YORP effect is extremely sensitive to surface details, like surface roughness, boulders and craters. A single large feature can alter YORP by tens  percent or more (Statler, 2009).

Given that the topographical features of asteroids are known in detail only in few cases, it is extremely difficult to predict the spin rate evolution on individual cases or on a population of homogeneous bodies (Statler, 2009). From this point of view a more statistical approach can be useful to estimate to which size the YORP effect has significantly altered the asteroids' rotation.

Now we derive a simple analytical scaling law for YORP effect, capable to compare the non-gravitational effects for two populations of bodies, not belonging to families, with different mean physical properties.

The infinitesimal force for YORP on a $dS$ element of the asteroid surface at temperature $T$, is given by (Spitale and Greenberg, 2001; Bottke et al., 2002; Bottke et al., 2006):

$$d\vec{F} = -\frac{2}{3}\frac{\varepsilon_t \sigma}{c} T^4 d\vec{S} \qquad (2)$$

In Equation (2), the Lambert emission model is assumed. The infinitesimal vector $d\vec{S}$ is orthogonal to the surface, $\varepsilon_t$ is the emissivity, $\sigma$ the Stefan-Boltzmann constant and $c$ the light speed in vacuum. The emissivity is dimensionless, being the ratio between the energy radiated by the material and the energy which would be radiated by a blackbody of equal effective temperature. So for a blackbody $\varepsilon_t = 1$, while for a real body $\varepsilon_t < 1$. Introducing the asteroid Bond albedo $A$ (or spherical albedo, defined as the ratio between the reflected flux and the incident flux), and the solar constant $F_S$ on asteroid surface, using energy conservation we can write (in the limit of zero thermal conductivity):

$$\varepsilon_t \sigma T^4 = F_S (1-A)\cos(z_S) = F_E \left(\frac{a_E}{R}\right)^2 (1-A)\cos(z_S) \qquad (3)$$

In Equation (3), $z_S$ is the zenith distance of the Sun, $R$ is the asteroid heliocentric distance, $F_E = 1378$ W m$^{-2}$ is the solar constant at Earth distance, and $a_E = 1$ AU is the mean distance Earth-Sun. Taking this into account, Eq. (2) can be rewritten as:

$$d\vec{F} = -\frac{2}{3}\frac{F_E}{c}\left(\frac{a_E}{R}\right)^2 (1-A)\cos(z_S) d\vec{S} \qquad (4)$$

The torque due to the infinitesimal YORP effect on a surface $dS$ at a $\vec{r}$ distance from the asteroid centre of mass is $d\vec{\tau} = \vec{r} \times d\vec{F}$. The total momentum is given by integrating over the whole asteroid surface. If the asteroid is perfectly spherical, the vectors $\vec{r}$ and $d\vec{S}$ are always parallel, so the torque is zero and total YORP effect is zero (Rubincam, 2000). Therefore, the non-zero contribution to the YORP effect is due to the body surface that has an irregular shape. Supposing that the asteroid has a spherical shape (with a mean diameter equal to $D$), but with asymmetric roughness on a total mean area equal to $S$. From Eq. (4), the order of magnitude of the torque resulting from the YORP effect will be given by:

$$\tau \propto -\frac{F_E}{c}\left(\frac{a_E}{R}\right)^2 (1-A)DS \tag{5}$$

In Equation (5) the value of the torque is zero when the asteroid is perfectly symmetric (ie $S = 0$). Now, the asteroid moment of inertia $I$, is proportional to:

$$I \propto MD^2 \propto \rho D^5 \tag{6}$$

In Equation (6), $M$ is the asteroid mass, while $\rho$ is the bulk density. Considered that the angular acceleration $\alpha$ is given by $\tau/I$, the angular acceleration due to YORP effect is proportional to:

$$\alpha \propto -f\frac{(1-A)}{\rho R^2 D^2} \tag{7}$$

In Equation (7), $f = S/D^2$ is the asteroid surface asymmetry factor. As we can see, the angular acceleration for YORP effect does not depend on the asteroid rotation frequency, thus any concentration in the spin distribution tends to be dispersed. From Equation (7) it follows that if two different asteroids (call them S and C) at a different distance from the Sun, are affected by the same YORP effect, we must respect the condition:

$$\frac{\rho_S R_S^2 D_S^2}{f_S(1-A_S)} \approx \frac{\rho_C R_C^2 D_C^2}{f_C(1-A_C)} \tag{8}$$

hence:

$$\left(\frac{D_C}{D_S}\right) \approx \left(\frac{R_S}{R_C}\right)\sqrt{\frac{(1-A_C)}{(1-A_S)}\left(\frac{f_C}{f_S}\right)\left(\frac{\rho_S}{\rho_C}\right)} \tag{9}$$

To summarize, in Eq. (9) $\rho$ is the mean asteroid bulk density, $D$ is the asteroid mean diameter, $R$ is the mean heliocentric distance and $f=S/D^2$ (where $S$ is the mean surface irregularity extension), is the asteroid surface asymmetry factor. Equation (9) gives an estimate of the diameter ratio that an S and C-type asteroid has to have if it is to be affected by the same YORP effect. It is interesting to note that, from Eq. (9), if we know the ratio between the transition diameters, we can derive the mean bulk density ratio for two different asteroid samples subjects to the YORP effect and that the knowledge of the mean bulk density of a population, allows the estimate of the other.

## 4 A first raw test for the YORP model

With the simple scaling law for YORP effect given by Equation (9) it is possible to justify the ratio between the transition diameters to the Maxwellian distribution for the samples of C and S asteroids. First, we need to calculate the Bond albedo from geometric albedo using the relationship (Bowell et al., 1989):

$$A = qp_v \tag{10}$$

In Equation (10), $q$ is the phase integral, while $p_v$ is the geometric albedo. Following Bowell et al., (1989), there is a relation between the phase integral and the slope parameter $G$ of the $H$-$G$ magnitude system:

$$q = 0.290 + 0.684\,G \tag{11}$$

For C asteroids the mean $G$ value is 0.07 ± 0.01, whereas for S is 0.24 ± 0.01 (Shevchenko and Lupishko, 1998). With these values, using Eq. (11) and the previous values for the geometric albedo, we have the following Bond albedo: $A_C$ = 0.02 ± 0.01 and $A_S$ = 0.09 ± 0.03, for C and S respectively. With respect to other quantities appearing in Eq. (9), for S we put $R_S$ = 2.538 ± 0.011 AU, while for C is $R_C$ = 3.013 ± 0.014 AU.

As mentioned in the introduction, for a first test of the model, we need to know a density ratio to ensure that we obtain a reasonable value of the ratio between the transition diameters, so we put $\rho_S/\rho_C \approx$ 2.0 ± 0.4, the mean value that comes from the values of Table 1. Moreover, for simplicity sake, we take $f_C/f_S = 1$, i.e. assuming that the two asteroids have the same asymmetry surface factor. With these values, from Eq. (9), we found:

$$\left(\frac{D_C}{D_S}\right) \approx 1.2 \pm 0.2 \tag{12}$$

This raw estimate of the ratio from our YORP model reasonably in agreement with the value of $1.5 \pm 0.1$ estimated for the samples of S and C asteroids, indeed $1.5\text{-}1.2 \approx 0.3 \pm 0.3$. It should also be noted that the used density ratio is an average value of plausible estimates but that the real value is still very uncertain (see Table 1). So, after this first raw test of the model, it is reasonable to assume that the ratio value between different diameter transitions to the Maxwellian distribution is due to the YORP effect.

## 5 More results from the YORP model

In this section two results from the YORP model are achieved: we provide a new average bulk density ratio between the two samples of C and S asteroids and show that our model proves the fact that the absolute magnitude of the asteroid near the transition diameter is approximately the same. Inverting Eq. (9), using the same values previously used in section 4 for the Bond albedo and heliocentric distance, and using the transition diameters ratio derived in section 2, we have:

$$\frac{\rho_S}{\rho_C} \approx 2.9 \pm 0.3 \tag{13}$$

This value represents a new estimate of the bulk density ratio between S and C asteroids and it appears physically reasonable. The indicative trend of the transition diameter for C and S asteroids on the whole Main Belt, calculated using Eq. (13), is shown in Fig. 6.

Based on this raw estimate, and by referring to Table 1, C-type asteroids appear less dense, or S-type asteroids are more dense, than estimated until now. Unfortunately, the YORP model does not allow individual density estimates. However, if we analyze the bulk densities of Table 1, we can see that the relative error in the average bulk density of S-type is 3.3% only, while for C-type it rises to 21%. The greater scattering for C asteroids is due to the fact that, being on average more distant from the Sun, it is more difficult to evaluate their gravitational effects over which the bulk density estimates of Table 1 are based.

So, from Table 1, taking as reference value the average bulk density of S asteroids, 2.72 ± 0.09 g cm⁻³, Eq. (13) gives us an average bulk density for C equal to 0.9 ± 0.1 g cm⁻³, a lower value than any other in Table 1. This result should be assumed with caution, because it is strictly model-dependent (e.g., for the assumption of the same asymmetry surface factor for both asteroid groups or if YORP is not the only process affecting the spin). Anyway, this indication of a low bulk density could be due to a high water ice percentage within the C-type asteroids. The recent discovery of water ice on the C-type asteroid (24) Themis ($a$ = 3.13 UA from the Sun, diameter of about 200 km), made by means of NASA's Infrared Telescope Facility, agrees with this interpretation (Rivkin and Emery, 2010). Of course, porosity makes it difficult to establish a direct relationship between the mean bulk density and the mean chemical composition of an asteroids sample, especially for smaller asteroids where porosity tends to increase (Baer and Chesley, 2008).

As we have seen in section 2, the best estimate of the transition diameter to the Maxwellian distribution for S-type asteroids is about 30 km, while for C-type asteroids is about 44 km. If we calculate the respective visual absolute magnitudes $H_V$, using the standard relation (Fowler and Chillemi, 1986; Warner et al., 2009):

$$D = \frac{1329}{\sqrt{p_V}} 10^{-\frac{H_V}{5}} \tag{14}$$

replacing the geometric albedo with the Bond albedo (Eq. 10 and 11), we find:

$$\begin{cases} H_S = 14.76 - 2.5 \log_{10}\left(D_S^2 A_S\right) \\ H_C = 14.44 - 2.5 \log_{10}\left(D_C^2 A_C\right) \end{cases} \tag{15}$$

Here, subscripts S and C mean S-type and C-type asteroids, while the diameter $D$ is in km. Substituting the above values, i.e. $A_S$ = 0.09, $D_S$ = 30 km, $A_C$ = 0.02 and $D_C$ = 44 km, in Eq. (15), we find $H_S$ = 10.0 mag and $H_C$ = 10.5 mag, hence $H_C$ - $H_S$ = 0.5 mag. Thus the transition occurs at about the same absolute magnitude, with a difference of only 0.5 mag.

Could this similarity be due to an observational bias? It is not likely since in paper I it has been shown that the periods of C-type and S-type asteroids are complete up to about 20 km in diameter ($H \approx 11$). Moreover, the YORP model can justify the small magnitude difference. Indeed, from Eq. (9), we can write:

$$D_C^2 A_C \approx 0.5\, D_S^2 A_S \qquad\qquad\qquad\qquad\qquad\qquad\qquad\qquad (16)$$

In Eq. (16), the 0.5 value comes from the density ratio given by Eq. (13). So, from Eq. (15), we obtain:

$$H_C - H_S = -0.32 - 2.5 \log_{10}\left(\frac{D_C^2 A_C}{D_S^2 A_S}\right) \approx -0.32 - 2.5 \log_{10}(0.5) \approx 0.4 \quad \text{mag} \qquad (17)$$

As we can see, the small magnitude difference between the transition diameters for C and S asteroids is perfectly justified by the YORP model.

To close this section, we will examine an unusual case in the MBAs, the Koronis family. The most prominent members of this family are of the S-type, have a semimajor axis between 2.83-2.95 AU, diameters ranging from 24 to 42 km and the estimated age is about 2-3 billion years, younger than the age of the Solar System. As asteroid collisions suggest, the Koronis family members should have spin rates that follow a Maxwellian distribution and random spin-axis orientation. In fact, it was found that the prograde rotators of this family have nearly the same rotation periods and the spin-axis are parallel in the space (Slivan, 2002). These peculiar features of the Koronis family could be explained with the YORP effect and spin-orbit resonance (Vokrouhlicky et al., 2003).

Our simple model predicts that at a distance near 3 AU from the Sun, the transition diameter for S-type asteroids is about 26 km (see Fig. 6), less than the diameter of the largest members of the Koronis family. This discrepancy may be due at least to two causes.

First, the model does not apply to few asteroids or bodies belonging to families (which were deleted from the database used, see introduction), second it is reasonable to expect that the surface of a recently formed body may have a greater number of small-scale irregularities than an older one. The small-scale irregularities may be due, for example, to the fallout of smaller fragments (e.g. boulders) on the surface of larger bodies of the family, after the collision from which it originates. Hereafter, due to continuous impacts with micrometeoroids and interplanetary dust grains, these structures tend to decay and disappear. So, according to Statler (2009), in the early billions of years of their life, the members of families are potentially more affected by the YORP than are asteroids that do not belong to families and it is reasonable to expect a greater transition diameter achieved in a short time interval. Indeed, Vokrouhlicky et al. (2003), found that an Ida shape model derived from Galileo spacecraft images evolves twice as fast as a more-rounded shape derived from lightcurve-inversion techniques.

If we compare MBAs S-type asteroids to those of the Koronis family, from Eq. (9) the ratio between the asymmetry surface factors is (the superscript letter K stands for Koronis):

$$\frac{f_s^K}{f_s} \approx \left( \frac{R_s^K D_s^K}{R_s D_s} \right)^2 \approx 2.7 \tag{18}$$

This ratio value seems physically accettable. As the decaying process goes on along with the gradual smoothing of the asteroid surface, the normal collisional evolution may regain importance, thereby slightly reducing the transition diameter value. Of course, at this stage, this is merely conjectural . We have not developed a collision-YORP evolution model to verify it.

## 6  Conclusions

Statistically, analysing two samples of C and S asteroids having a known rotation period, it was found that the transition diameters to the Maxwellian distribution are different,  $30 \pm 1$ km and $44 \pm 2$ km respectively. This ratio between the transition diameters $(1.5 \pm 0.1)$ can be explained as a consequence of the YORP effect on the two asteroids samples which differ in their average distance from the Sun, bulk density and albedo, although other physical causes can not be completely ruled out. A simple theoretical model YORP -based provides, for the transition diameter ratio, the value of  $1.2 \pm 0.2$, compatible with the observed value.

The YORP model developed here can also be used to estimate a ratio between the bulk densities of two different asteroids samples. For the sample of S and C asteroids studied here, we found $\rho_S / \rho_C \approx 2.9 \pm 0.3$, a value which appears physically reasonable. Moreover, from Table 1, taking as reference the less uncertain value for the average bulk density of S-type asteroids, $2.72 \pm 0.09$ g cm$^{-3}$, it is possible to estimate an average bulk density for C-type equal to $0.9 \pm 0.1$ g cm$^{-3}$. This result should be taken with caution, but it is probably due to a high water ice percentage within the C-type asteroids sample; however, there is a chance that also  porosity  plays a major role, and the problem is still open. The recent discovery of water ice on the C-type asteroid (24) Themis seems to conform with this interpretation but more data are necessary.

 Finally, the model is also suitable to justify the little difference in absolute magnitude for asteroids C and S near the transition diameter.

Of course, even if we have done our best to consider all uncertainties, the results put forward  in this paper may change in the event that the asteroid's uncertainty on diameters and rotation periods are reduced and the taxonomic classification is improved.

Finally, it should be noted that the proposed explanation (e.g. YORP) for the difference of the transition diameters is a plausible hypothesis consistent with the data, but it needs more in depth attention and research work .

Besides, it shouldn't be overinterpreted with the current limited understanding of the physical properties of asteroids. For these reasons, repeated and more advanced analysis will be necessary in years to come.

**Acknowledgments**

A special thanks to Peter Pravec (Astronomical Institute, Czech Republic) for the very constructive comments on the manuscript, and to Alan W. Harris (Space Science Institute) and Brian D. Warner (Palmer Divide Observatory) for the quick assistance with the new ALCDB version. Thanks to Paolo Tanga (Observatoire de la Côte d'Azur), Stefano Mottola (DLR - German Aerospace Center), and Mario di Martino (Torino Astronomical Observatory) for critically reading the manuscript and to Giovanni De Sanctis (Torino Astronomical Observatory), for his help in obtaining references . Thanks to the anonymous referees that with their advice have contributed to the improvement of the work.

Research on asteroids at OAVdA is supported by Director, Enzo Bertolini, and funded with a European Social Fund grant from the Regional Government of the Regione Autonoma della Valle d'Aosta (Italy). This work has made use of NASA's Astrophysics Data System Bibliographic Services.

**Appendix**

**Estimate of the minimum $n$ value to explore asteroids ' diameter-rotation frequency space, and of diameter transition uncertainty.**

In this appendix we estimate the minimum number of virtual diameters and rotation frequency sets, that is necessary to simulate to properly explore the entire space within $s$ sigma (where $s = 1, 2, 3$). In general, any virtual diameters and rotation frequencies set can be written as:

$$\vec{S} = \left( D_1 + e_1, D_2 + e_2, \dots, D_m + e_m, F_1 + \delta_1, \dots, F_k + \delta_k \right) \tag{A.1}$$

Where $D_i$, with $i = 1, 2 \dots m$, are the diameters observed for $m$ asteroids, while the $e_i$ values are the associated uncertainties with normal distribution (if systematic errors are negligible). Similarly, $F_i$ is the rotation frequency U=2, with $i = 1, 2…k$ (usually $k \leq m$), and $\delta_i$ the associated uncertainties. In our model, the normal distribution has a mean value equal to $D_i$ (or $F_i$) and standard deviation $\sigma_i = 0.2 \cdot D_i$ (or $\sigma_i = 0.2 \cdot F_i$). In the $m+k$-dimensional space, the volume occupied by all asteroids, with their uncertainties within $s\sigma$, is given by:

$$V_{m+k} = 2s \prod_{i=1}^{m+k} \sigma_i \qquad \text{(A.2)}$$

The volume in the diameters and rotation frequency space, explored with the creation of $n$ virtual asteroid sets obtained by adding normal noise to the observed set, can be approximately measured as follows:

$$V_{m+k}^n \approx \prod_{i=1}^{m} \left( \max\left(D_i^n\right) - \min\left(D_i^n\right) \right) \prod_{i=1}^{k} \left( \max\left(F_i^n\right) - \min\left(F_i^n\right) \right) \qquad \text{(A.3)}$$

Where $D_i^n$ is the $n$-tuple of virtual diameters and $F_i^n$ is the $n$-tuple of virtual rotation frequencies, obtained for the same asteroid, with the simulation of $n$ virtual sets. In this way, the explored fraction of diameters and rotation frequencies space is given by:

$$Q_n = \frac{V_{m+k}^n}{V_{m+k}} \qquad \text{(A.4)}$$

Performing some simulations with $s = 3$, we find a $Q_n$ value close to 1 with $n \approx 450$ for C-type asteroids and with $n \approx 250$ for S-type asteroids. These ones are the $n$ values adopted in the paper.

As explained in section 2 of this paper, for each inferior value diameter of the virtual asteroids sample, Kuiper test is carried out. If the rotation frequency distribution is consistent with a Maxwellian, at least with a 5% confidence level, the value "0" is assigned, otherwise "1". Thus, in the scale of increasing diameters, at 1 km step, the passage from non-Maxwellian to Maxwellian distribution is indicated by the transition from 1 to 0 in a string of digits like this: 1-1-1-1-1-0-0-0-0-0. We call this type of string "step sequence". However, in the sequence of 0 and 1, a first transition, a return to non-compatibility, a new transition and so on , may happen. For an example of this case let us imagine a string of this type: 1-1-1-1-1-0-1-0-0-1-0-0-0-0-0. We call this type of string "many steps sequence". In this case it becomes more difficult to understand where the transition diameter to Maxwellian exactly falls.

In the simulations done, the diameter transition was searched within the range 20-70 km for C-type and in the range 15-50 km for S-type.

If, as a transition diameter, we consider the mean diameter value between "1" and "0" in the first appearance of the sequence 1-0, the results are the following (the mean of 10 independent runs):

$D_C = 44.7 \pm 0.8$ km

$D_S = 29.3 \pm 0.4$ km

To determine the uncertainty due to many steps sequences on the simple step sequences, we have also considered the transitions diameters given by the sequences 1-0-0 and 1-0-0-0, i.e. where the Maxwellian character of the distribution remains for one or two times after the first transition. In this case we have the following results (mean of 10 independent runs):

$D_C^{1-0-0} = 44.1 \pm 1.1$ km

$D_C^{1-0-0-0} = 41.9 \pm 0.9$ km

$D_S^{1-0-0} = 29.6 \pm 0.4$ km

$D_S^{1-0-0-0} = 30.0 \pm 0.4$ km

As we can see, the transition diameter for C-type decreases by 2.8 km, whereas the one of S-type remains substantially unchanged. On the cautious basis of these results, it seems reasonable (without underestimating uncertainties) to assume these values for the transition diameters:

$D_C = 44 \pm 2$ km

$D_S = 30 \pm 1$ km

These are the values given in the text.

The dataset is available on http://www.minorplanetobserver.com/astlc/LightcurveParameters.htm.

Table 1

Dynamical bulk density estimate for C and S Main Belt asteroid (Krasinsky et al., 2002). The bulk density values are in g cm$^{-3}$. For the physical meaning of solutions 2, 3 and 4 see the paper mentioned above. The last column shows the results derived from the YORP model discussed in the text.

|  | Solution 2 | Solution 3 | Solution 4 | YORP model |
|---|---|---|---|---|
| Largest asteroids number | 300 | 351 | 351 | 481 |
| $\rho_S$ | 2.71 ± 0.01 | 2.82 ± 0.03 | 2.64 ± 0.08 | --- |
| $\rho_C$ | 1.34 ± 0.04 | 1.12 ± 0.04 | 1.74 ± 0.11 | 0.9 ± 0.1 |
| $\rho_S/\rho_C$ | 2.02 ± 0.06 | 2.52 ± 0.09 | 1.52 ± 0.11 | 2.9 ± 0.3 |

Table 2

List of C and S asteroids discarded around Maxwellian transition diameter.

| S-type asteroids discarded (20-50 km range) | |
|---|---|
| Motivation | Asteroids progressive numbers |
| Probably C-type for $p_v$ value | (829), (1159), (1233), (1244) |
| C-type | (723) |
| X-type | (1407) |
| C-type asteroids discarded (30-60 km range) | |
| Motivation | Asteroids progressive numbers |
| Probably S-type for $p_v$ value | (297), (430), (655), (894), (987), (1113), (1116), (1232), (1276), (1481), (1572) |
| X-type | (1039), (1323), (1428), (1502) |
| No physical data | (942), (1125), (1157), (1175), (1319), (1488), (1910), (2104), (2193), (2288), (2323), (2347), (2612), (2802), (3015), (3106), (3259), (3300), (3431), (3514), (3557), (3761), (3843), (4045), (5357), (6393), (12559) |

Table 3

Fractions of asteroids with diameters from SIMPS vs. lower diameter limit.

| Full sample of S-type asteroids | Full sample of C-type asteroids | Inferior diameter value (km) |
|---|---|---|
| 0.35 | 0.68 | 1 |
| 0.60 | 0.73 | 10 |
| 0.95 | 0.96 | 50 |
| 0.95 | 0.98 | 100 |

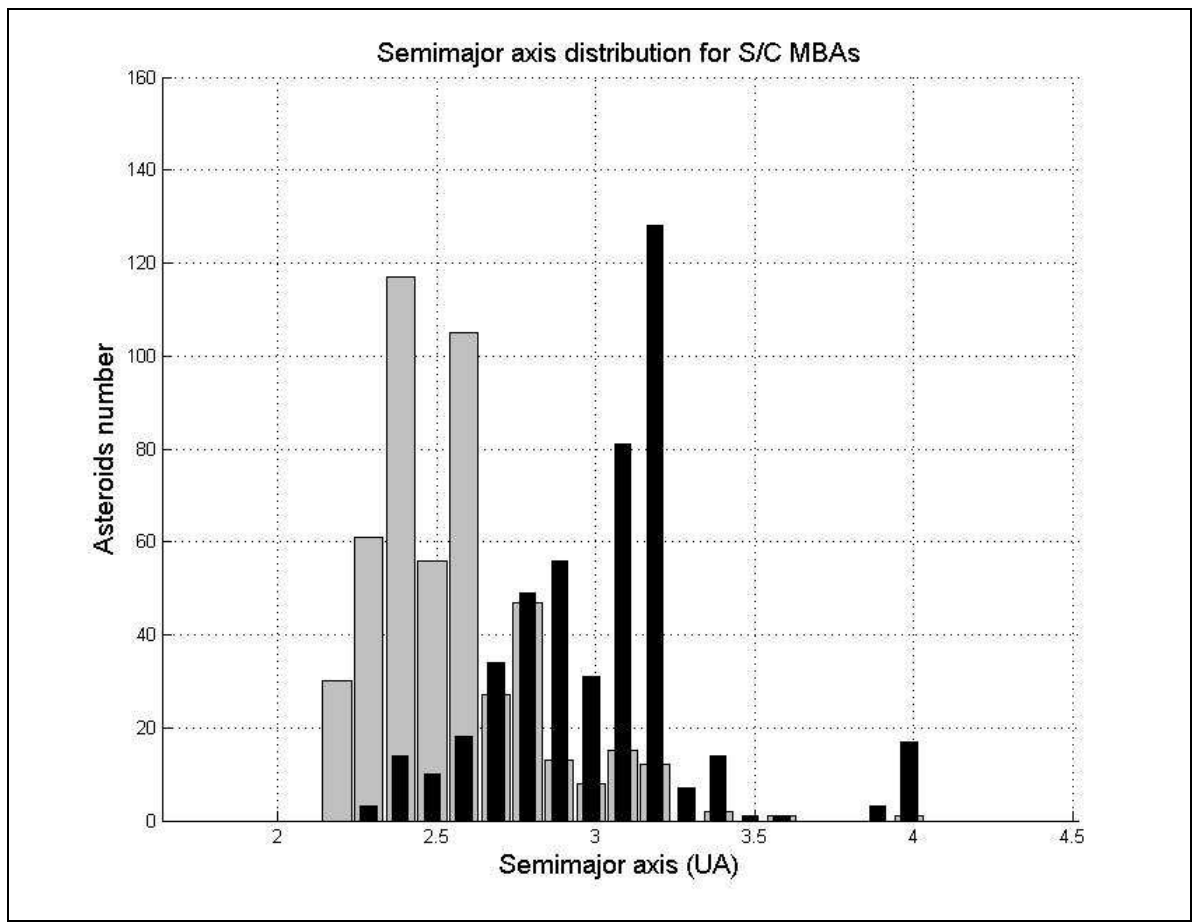

Fig. 1 Comparison between the semimajor axis distributions for C (black bar) and S (grey bar) MBAs of the samples. The distribution dichotomy between the two samples is evident.

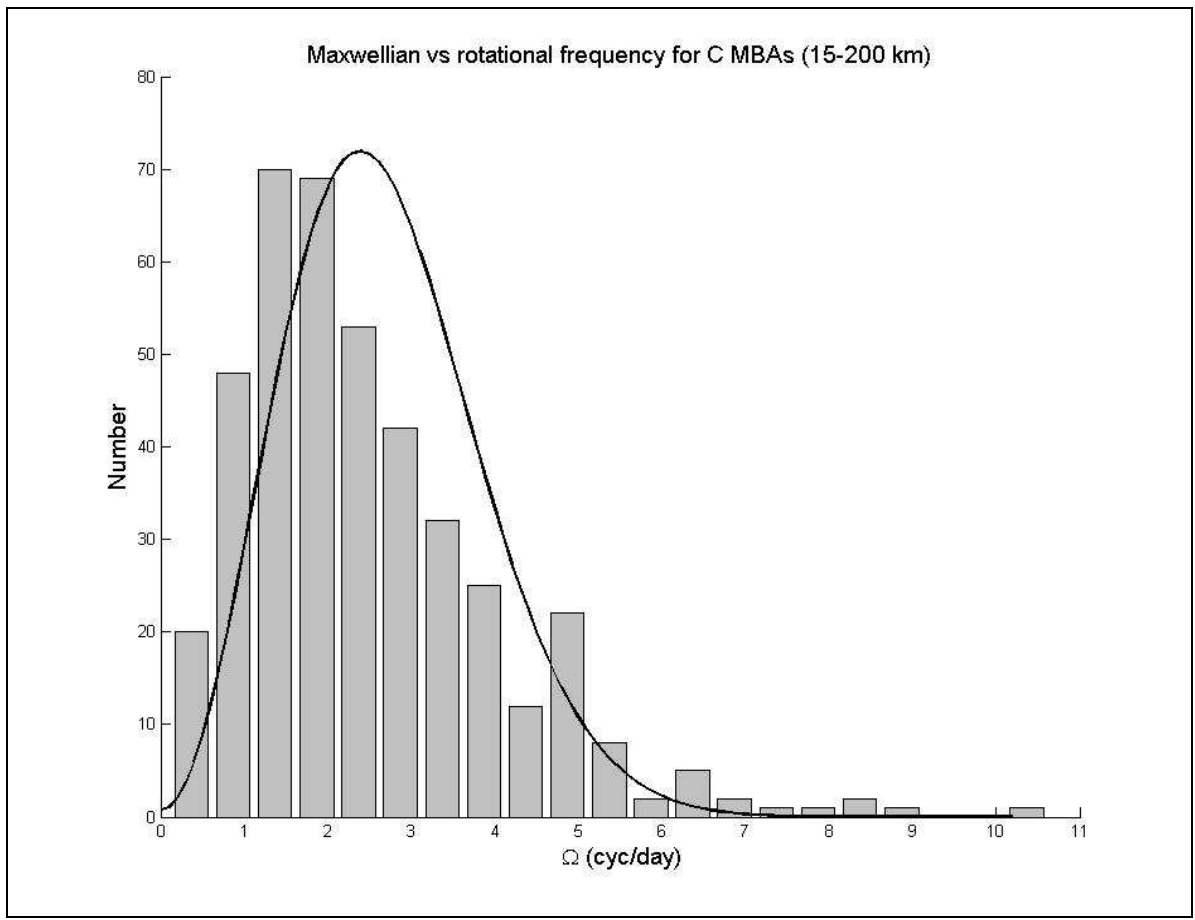

Fig. 2 The three-dimensional Maxwellian distribution vs the rotation frequency distribution (not normalized) for the sample of 415 C-type MBAs in the range 15-200 km. There is an excess of slow rotators, a lack of medium rotators and an excess of fast rotators.

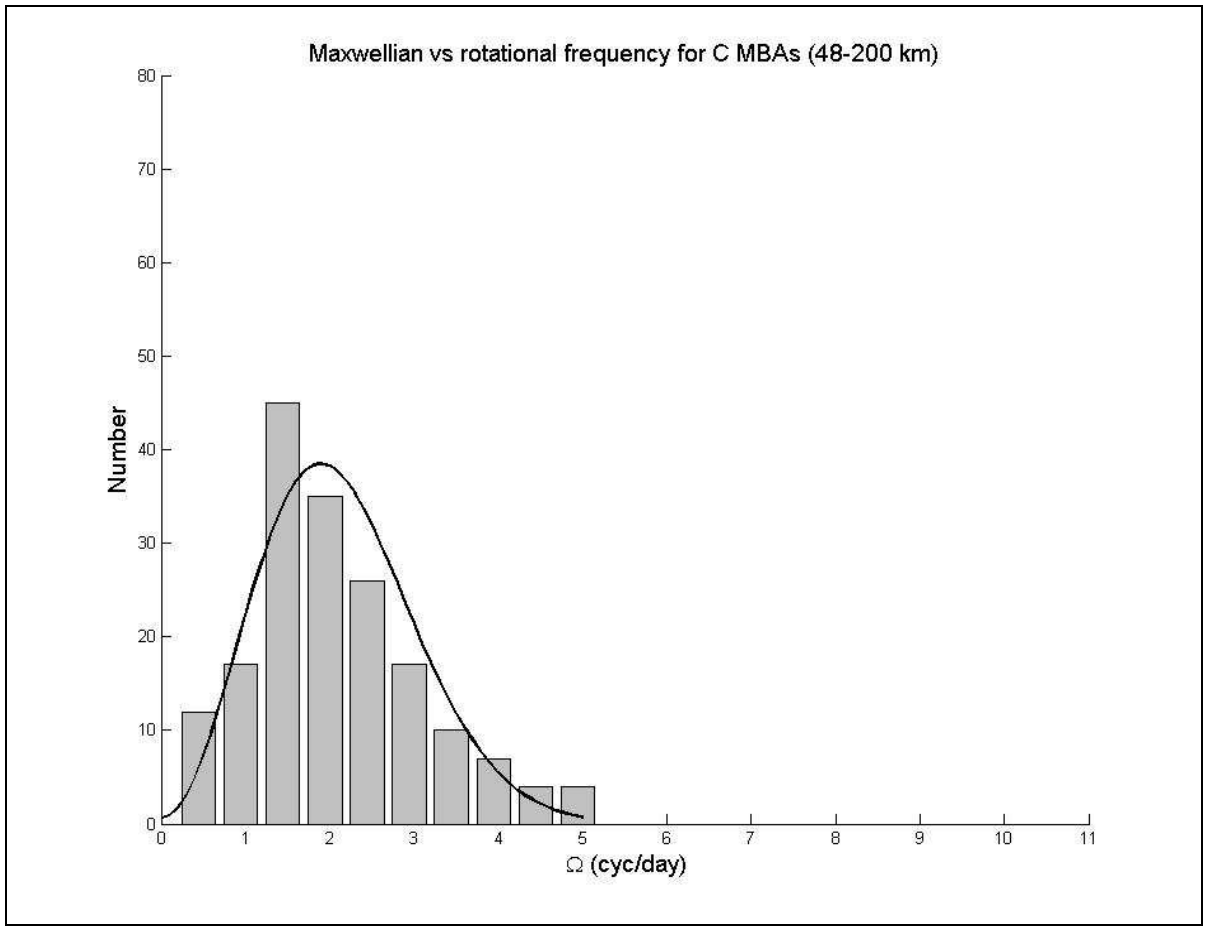

Fig. 3 The three-dimensional Maxwellian distribution vs the rotation frequency distribution (not normalized) for the sample of 176 C-type MBAs in the range 48-200 km. The slow/fast rotators excess is considerably reduced respect to Fig. 2.

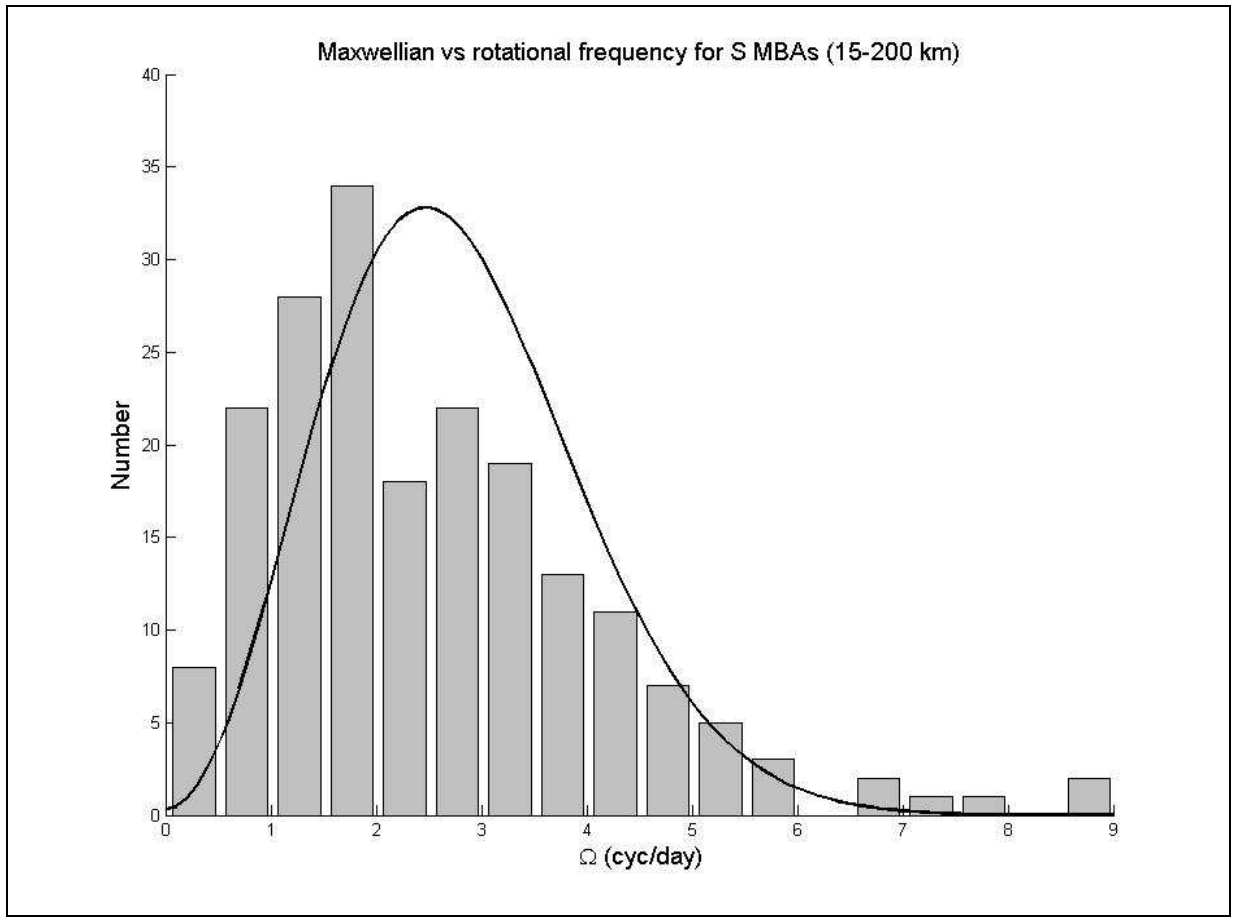

Fig. 4 The three-dimensional Maxwellian distribution vs the rotation frequency distribution (not normalized) for the sample of 194 S-type MBAs in the range 15-200 km. As for C asteroids, there is an excess of slow rotators, a lack of medium rotators and an excess of fast rotators.

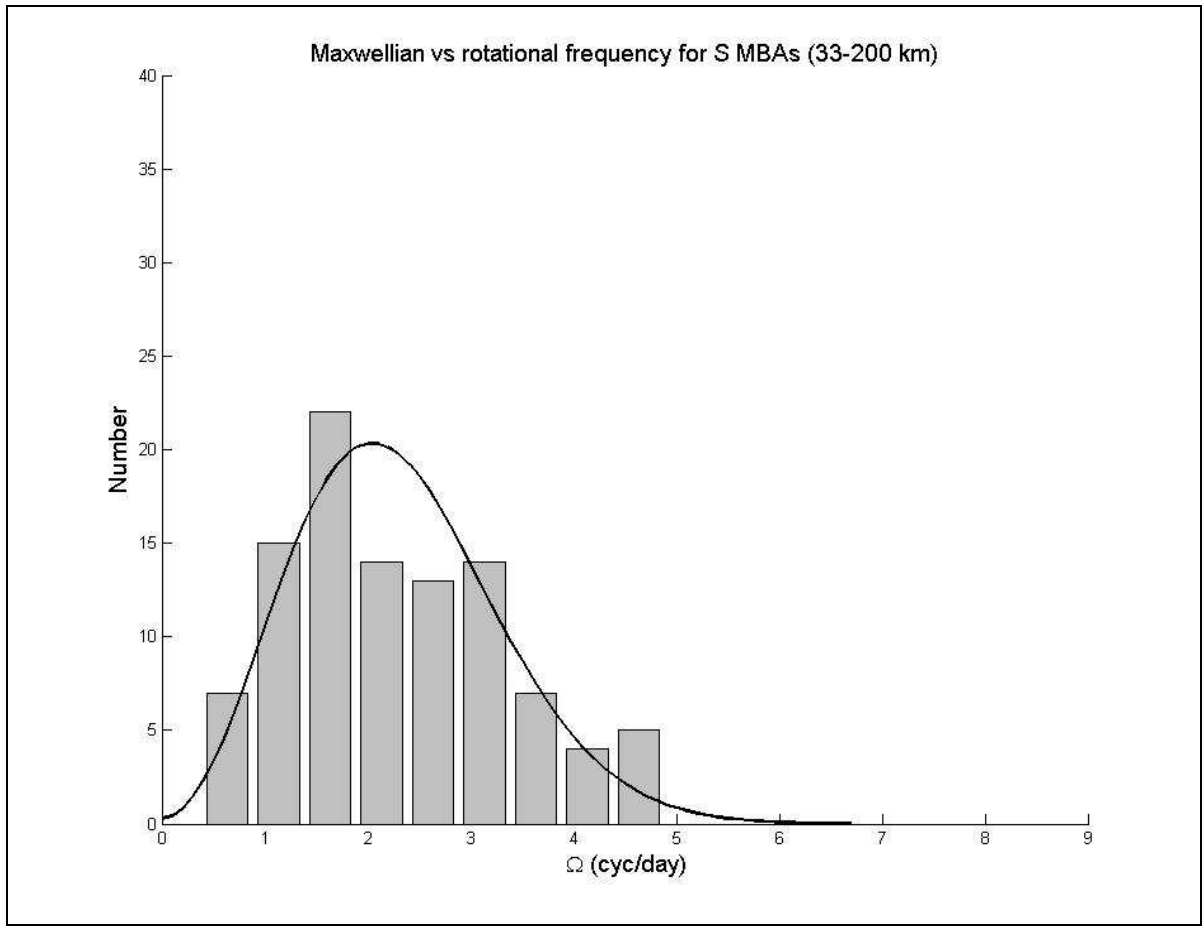

Fig. 5 The three-dimensional Maxwellian distribution vs the rotation frequency distribution (not normalized) for the sample of 99 S-type MBAs in the range 33-200 km. The slow/fast rotators excess is considerably reduced respect to Fig. 4.

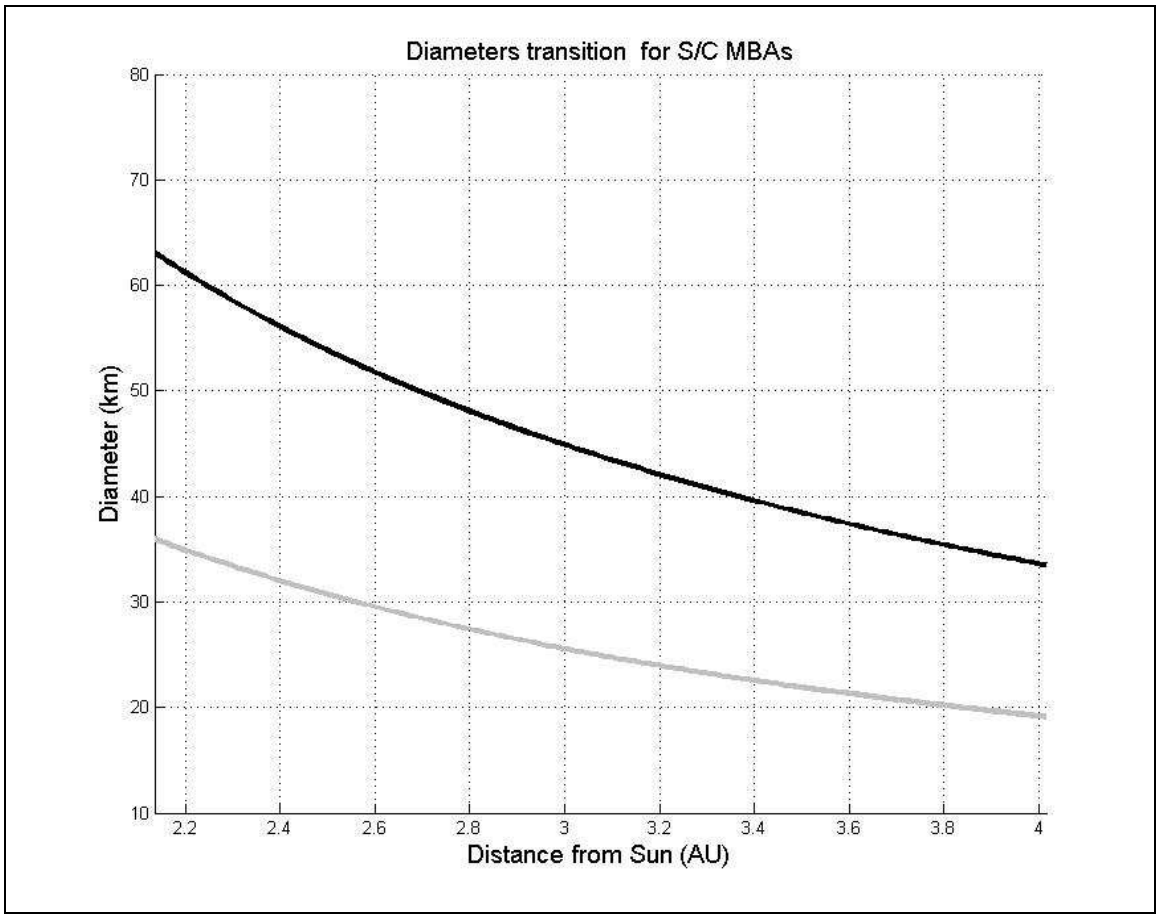

Fig. 6 Estimated diameter transition to Maxwellian distribution, from Eq. (9), for C (black line) and S (grey line) MBAs. The density ratio adopted here is from Eq. (13).